\documentclass[twocolumn,showpacs]{revtex4}
\usepackage{epsfig}
\begin{document}

\title{Entanglement of distant electron interference experiments}
\author{D.I. Tsomokos, C.C. Chong, A. Vourdas}
\affiliation{Department of Computing, School of Informatics,
 University of Bradford, Bradford BD7 1DP, United Kingdom}

\begin{abstract}
Two electron interference experiments which are far from each other, are
considered. They are irradiated with correlated nonclassical electromagnetic
fields, produced by the same source. The phase factors are in this case
operators, and their expectation values with respect to the density matrix of
the electromagnetic field quantify the observed electron fringes. The
correlated photons create correlations between the observed electron
intensities. Both cases of classically correlated (separable) and quantum
mechanically correlated (entangled) electromagnetic fields are considered. It
is shown that the induced correlation between the distant electron
interferences is sensitive to the nature of the correlation between the
irradiating photons.
\end{abstract}

\emph{Phys. Rev.} \textbf{A 69}, 013810 (2004).

\pacs{42.50.Dv; 85.35.Ds; 73.23.-b} \maketitle

\section{Introduction}
Interference of electrons encircling a magnetostatic flux has been studied
extensively since the work of Aharonov and Bohm \cite{AB,SM}. These ideas have
been applied in various contexts, for example, in magnetoconductance
oscillations in mesoscopic rings \cite{SS}, ``which-path'' experiments
\cite{WP}, and neutron interferometry \cite{NI}.

The Aharonov-Bohm effect can be generalized by replacing the magnetostatic
flux with an electromagnetic field. The objective in this ``ac Aharonov-Bohm''
effect is very different from the ``dc Aharonov-Bohm'' effect (with
magnetostatic flux). In the latter case the physical reality of the vector
potential has been demonstrated and the subtleties of quantum mechanics in
nontrivial topologies have been studied. The former case constitutes a
nonlinear device, where the interaction between the interfering electrons and
the photons leads to interesting nonlinear phenomena. Indeed the nonlinearity
can be seen in the intensity of the interfering electrons which is a
sinusoidal function of the time-dependent magnetic flux. In  Refs. \cite{1,2}
the interference of electric charges in the presence of both classical and
nonclassical electromagnetic fields, has been studied. It has been shown that
the quantum noise of the electromagnetic field affects the phase factor.

In this paper we consider two Aharonov-Bohm interference devices which are far
from each other. Each of them is irradiated with a nonclassical
electromagnetic field. The aim of the paper is to consider entanglement
between the two electromagnetic modes irradiating the two Aharonov-Bohm
interference devices and study the correlations between the interfering
electrons in the two devices. We note that in Refs. \cite{2} we have studied
the effect of photon entanglement on a single Aharonov-Bohm interference
device. Here we consider two electron interference devices far apart from each
other, and show that the two electron interferences are correlated due to the
entanglement between the two electromagnetic fields.

In Sec. II we describe the experiment. We show that the joint electron
intensity depends on the density operator describing the two-mode
electromagnetic field. In Sec. III we consider two cases for the density
operator of the field, separable and entangled \cite{EN}. We conclude in Sec.
IV with a discussion of the results.

\section{Electron interference}

We consider two electron interference experiments far apart from each other,
which we refer to as A and B (Fig. 1). They are irradiated with
electromagnetic fields. Each electron beam splits into two paths
$C_{A0},C_{A1}$ and $C_{B0},C_{B1}$ (paths with higher winding numbers are
ignored).

Let $\phi_A$ be the time dependent flux threading the loop $C_{A1}-C_{A0}$.
The electron wave function at the point $x_A$ is given by \cite{1,2}
\begin{equation} \label{def_L}
\Psi_A(x_A)=\psi_{A0}(x_A)+\exp(ie\phi_A)\psi_{A1}(x_A),
\end{equation}
where $\psi_{A0}(x_A)$ and $\psi_{A1}(x_A)$ are the electron wave functions
associated with the paths $C_{A0}$ and $C_{A1}$, correspondingly. This leads
to the electron intensity
\begin{eqnarray} \label{I_device_A}
I_A(\sigma_A)&=&1+\cos(\sigma_A+e\phi_A), \\
\sigma_A &\equiv& \arg[\psi_{A1}(x_A)]-\arg[\psi_{A0}(x_A)],
\nonumber
\end{eqnarray}
for the case of equal splitting between the two wave functions
($|\psi_{A0}(x_A)|^2=1/2=|\psi_{A1}(x_A)|^2$). We note that the phase
difference $\sigma_A$ is effectively a rescaled position $x_A$ on the screen.
All the results below are in terms of $\sigma_A$ (and $\sigma_B$).

\begin{figure}[htbp]
\begin{center}
\scalebox{0.7}{\includegraphics{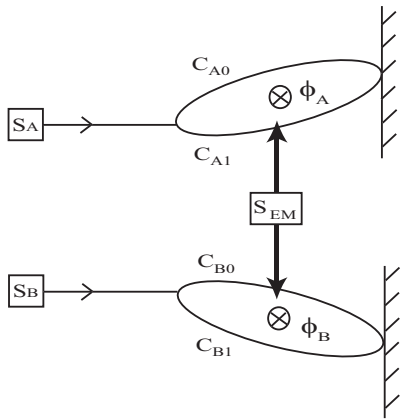}}
\end{center}
\caption{Two electron interference experiments which are far from each other
are irradiated with nonclassical electromagnetic fields. The two
electromagnetic fields in the two experiments are produced by the source
$S_{EM}$ and are correlated.}
\end{figure}

\subsection{Nonclassical electromagnetic fields}
As explained in our previous work \cite{1,2} for a nonclassical
electromagnetic field of frequency $\omega$ the flux $\phi$ is an operator.
The dual quantum variables of the electromagnetic field are the vector
potential $A_i$ and the electric field $E_i$. Although the dual quantum
variables are local quantities, we consider loops which are small in
comparison to the wavelength and after integration we get $\hat \phi=\oint_C
A_i dx_i$ and the electromotive force $\hat V=\oint_C E_i dx_i$ as dual
quantum variables. We next introduce the corresponding creation and
annihilation operators, $\hat a^{\dag}=2^{-1/2}\xi^{-1}(\hat\phi-i\omega^{-1}
\hat V)$ and $\hat{a}=2^{-1/2}\xi^{-1}(\hat{\phi}+i\omega^{-1} \hat{V})$,
where $\xi$ is a constant proportional to the area enclosed by the loop and in
units $k_B=\hbar=c=1$. We work in the ``external field approximation'' where
the back-reaction from the electrons on the electromagnetic field is
neglected. This is valid for external fields which are strong in comparison to
those produced dynamically by the electrons (back-reaction). In this case we
get $\hat \phi (t) =2^{-1/2}\xi [\exp(i\omega t)\hat{a}^\dagger +
\exp(-i\omega t)\hat{a}]$.

Exponentiation of the magnetic flux operator yields the phase factor
$\exp[ie\hat\phi(t)]$, which becomes
\begin{equation} \label{q_definition}
\exp[ie\hat \phi (t)]=D[iq\exp(i\omega t)]; \;\;\; q=2^{-1/2}\xi e,
\end{equation}
where  $D(x)=\exp(x\hat a^{\dag}-x^* \hat a)$ is the displacement operator. In
order to find expectation values we take the trace of the
$\exp[ie\hat\phi(t)]$ operator with respect to the density matrix $\rho$
describing the nonclassical electromagnetic field:
\begin{eqnarray} \label{W_phase_factor}
{\rm Tr} \{\rho \exp[ie\hat \phi (t)]\}
&=&\mbox{Tr}[\rho D(\lambda)]\equiv \tilde W (\lambda)\nonumber\\
\lambda&=&iq\exp(i\omega t).
\end{eqnarray}
Here $\tilde W$ is the Weyl or characteristic or ambiguity function
(\emph{cf.} Ref. \cite{WW} and references therein). The tilde in the notation
reflects the fact that the Weyl function is the two-dimensional Fourier
transform of the Wigner function (which is usually denoted by $W$).

\subsection{Correlated electron intensities}
Let $\rho$ be the density operator describing the two-mode nonclassical
electromagnetic field in both devices. The first mode of frequency $\omega _1$
interacts with electrons in experiment ${\rm A}$ and its density matrix is
$\rho_A\equiv {\rm Tr}_B \rho$. The second mode of frequency $\omega _2$
interacts with electrons in experiment ${\rm B}$ and its density matrix is
$\rho_B\equiv {\rm Tr}_A \rho$.

For nonclassical electromagnetic fields the flux and consequently the
intensity of Eq. (\ref{I_device_A}) are operators. In order to find the
expectation value of the intensity we calculate its trace with respect to the
appropriate density matrix and using Eq. (\ref{W_phase_factor}) we find
\begin{eqnarray} \label{I_Left_Right}
I_A(\sigma_A)&=& \mbox{Tr}\{\rho_A
[1+\cos(\sigma_A+e\hat\phi_A)]\}  \nonumber \\
&=&1 +
|\tilde{W}(\lambda_A)|\cos\{\sigma_A+\arg[\tilde{W}(\lambda_A)]\}.
\end{eqnarray}
where $\lambda_A=iq\exp(i\omega _1 t)$. As we have explained in detail in our
previous work \cite{1,2}, the visibility in this case is
$|\tilde{W}(\lambda_A)|$ which takes values less than $1$. It has been shown
there that this is intimately related to the quantum uncertainties in the
electric and magnetic fields and consequently the reduction of the visibility
from $1$ to $|\tilde{W}(\lambda_A)|$ is due to the quantum noise in the
nonclassical fields.

Similarly the electron intensity in experiment ${\rm B}$ is
\begin{eqnarray}
I_B(\sigma_B)&=& \mbox{Tr}\{\rho_B
[1+\cos(\sigma_B+e\hat\phi_B)]\}\nonumber\\
&=&1 +
|\tilde{W}(\lambda_B)|\cos\{\sigma_B+\arg[\tilde{W}(\lambda_B)]\},
\end{eqnarray}
where $\lambda_B =i q \exp (i \omega_2 t)$.

We next consider the joint electron intensity in the two experiments.  It is
given by
\begin{eqnarray} \label{quantum_I}
\lefteqn{I(\sigma_A,\sigma_B) = } \nonumber \\
&& \mbox{Tr}\{\rho [1+\cos(\sigma_A+e\hat\phi_A)][1+\cos(\sigma_B
+e\hat\phi_B)]\}.
\end{eqnarray}
The correlations between the electron interferences in the two experiments are
quantified with the ratio
\begin{eqnarray} \label{ratio}
R =\frac { I(\sigma_A ,\sigma_B )} {I_A (\sigma_A ) I_B (\sigma_B
)}.
\end{eqnarray}

\section{Examples}
Two mode density matrices are factorizable (uncorrelated) if they can be
written as $\rho=\rho_{A}\otimes\rho_B$. They are separable (classically
correlated) if they can be written as $\rho=\sum_i p_i \rho_{A,i}\otimes
\rho_{B,i}$ where $p_i$ are probabilities. Density matrices which cannot be
written in one of these two forms are entangled (quantum mechanically
correlated). There has been a lot of work on criteria which distinguish
separable and entangled states \cite{EN}. In this paper we compare and
contrast the influence of separable and entangled photon states on two distant
electron interference experiments.

We consider two cases for the density operator $\rho$ of the two-mode
electromagnetic fields. The first is the separable (classically correlated)
density matrix
\begin{equation} \label{rho_sep}
\rho_{{\rm sep}}= \frac{1}{2}(|01\rangle \langle 01| +|10\rangle \langle 10|).
\end{equation}
The second is the entangled state
$|\mbox{S}\rangle=2^{-1/2}(|01\rangle
+|10\rangle)$
with corresponding density matrix
\begin{eqnarray} \label{rho_ent}
\rho_{{\rm ent}}&=&\rho_{{\rm sep}}+\frac{1}{2}(|01\rangle \langle 10|
+|10\rangle \langle 01|),
\end{eqnarray}
where $\rho_{\rm sep}$ is given by Eq. (\ref{rho_sep}). The difference between
$\rho_{\rm sep}$ and $\rho_{\rm ent}$ lies in the above non-diagonal elements.

\subsection{Classically correlated number eigenstates}
In the case of separable electromagnetic fields of Eq. (\ref{rho_sep}) the
electron intensities are
\begin{eqnarray} \label{I_A}
I_A(\sigma_A)&=&1+\alpha\cos\sigma_A, \\
I_B(\sigma_B)&=&1+\alpha\cos\sigma_B, \nonumber
\end{eqnarray}
where
\begin{equation} \label{constants_A}
\alpha=\frac{2-q^2}{2}\exp\left(-\frac{q^2}{2}\right).
\end{equation}
As explained in detail in Refs. \cite{1,2} the visibility corresponding to
$I_A$ or $I_B$ is $\alpha<1$, due to the noise in the nonclassical
electromagnetic fields.

We also calculate the joint intensity
\begin{eqnarray} \label{I_sep}
I_{\rm sep}(\sigma_A,\sigma_B) &=& 1 + \alpha(\cos\sigma_A+\cos\sigma_B)\nonumber \\
& &+ 2\beta\cos\sigma_A\cos\sigma_B,
\end{eqnarray}
where
\begin{eqnarray} \label{constant_B}
\beta=\frac{1-q^2}{2}\exp\left(-q^2 \right).
\end{eqnarray}
The ratio $R$ of Eq. (\ref{ratio}) is
\begin{eqnarray} \label{R_sep}
\lefteqn{R_{\rm sep}(\sigma_A,\sigma_B)=} \nonumber \\
&&\frac{1+\alpha(\cos\sigma_A+\cos\sigma_B)+2\beta\cos\sigma_A\cos\sigma_B}
{(1+\alpha\cos\sigma_A)(1+\alpha\cos\sigma_B)}.
\end{eqnarray}
$R_{\rm sep}(\sigma_A,\sigma_B)$ is periodic in
$\sigma_A$ and $\sigma_B$ with period $2\pi$ for each of the screen
positions. Its stationary points are such that $\frac{\partial
R_{\rm sep}}{\partial\sigma_A}=0=\frac{\partial R_{\rm
sep}}{\partial\sigma_B}$ and it can easily be shown that
\begin{equation} \label{R_inequality}
\frac{1-2\alpha+2\beta}{(1-\alpha)^2} \leq R_{\rm sep}(\sigma_A,\sigma_B) \leq
\frac{1+2\alpha+2\beta}{(1+\alpha)^2}.
\end{equation}
The global minimum occurs at the point $(\sigma_A=\pi,
\sigma_B=\pi)$ and the global maxima at the points
$(\sigma_A=0~\mbox{or}~2\pi, \sigma_B=0~\mbox{or}~2\pi)$.

We note that for factorizable (uncorrelated) electromagnetic fields $R=1$. In
the example of separable (classically correlated) electromagnetic fields of
Eq. (\ref{rho_sep}) we get $R_{\rm sep}$ independent of time, which takes
values less than $1$ (the upper bound in the inequality (\ref{R_inequality})
is slightly less than $1$).

\subsection{Entangled number eigenstates}
We now consider the entangled electromagnetic fields of Eq. (\ref{rho_ent}).
In this case the electron intensities $I_A(\sigma_A)$ and $I_B(\sigma_B)$ are
the same as in Eq. (\ref{I_A}). The joint electron intensity is
\begin{eqnarray} \label{I_ent}
\lefteqn{I_{\rm ent}(\sigma_A,\sigma_B) =I_{\rm sep}(\sigma_A,\sigma_B)} \nonumber \\
&-& q^2\exp(-q^2) \sin\sigma_A \sin\sigma_B \cos[(\omega_1-\omega_2)t],
\end{eqnarray}
and the ratio of Eq. (\ref{ratio}) is
\begin{eqnarray} \label{R_ent}
R_{\rm ent}(\sigma_A,\sigma_B)&=&R_{\rm
sep}(\sigma_A,\sigma_B) -q^2\exp(-q^2) \nonumber \\
& &\times \frac{\sin\sigma_A \sin\sigma_B
\cos[(\omega_1-\omega_2)t]
}{(1+\alpha\cos\sigma_A)(1+\alpha\cos\sigma_B)}.
\end{eqnarray}
It is seen that $I_{\rm ent}(\sigma_A,\sigma_B)$ is equal to $I_{\rm
sep}(\sigma_A,\sigma_B)$ of Eq. (\ref{I_sep}) plus an extra term, which
oscillates in time with frequency $\omega _1-\omega _2$ around this value.
In the case $\omega_1 =\omega_2$ the electron intensity $I_{\rm ent}(\sigma_A,\sigma_B)$
differs from the electron intensity $I_{\rm sep}(\sigma_A,\sigma_B)$ by a constant
(which depends on $\sigma_A$, $\sigma_B$).
Similar comments apply to $R_{\rm sep}(\sigma_A,\sigma_B)$ and $R_{\rm ent}(\sigma_A,\sigma_B)$

\subsection{Numerical results}
In all numerical results the electromagnetic fields have frequencies
$\omega_1=1.2\times 10^{-4}$ and $\omega_2=10^{-4}$, and the parameter
$\xi=1$. Fig. 2 shows the $I_{\rm sep}(\sigma_A,\sigma_B)$ of Eq.
(\ref{I_sep}) as a function of $\sigma_A$ and $\sigma_B$. Fig. 3 shows the
$R_{\rm sep}(\sigma_A,\sigma_B)$ of Eq. (\ref{R_sep}). We note that in our
example the $R_{\rm sep}$ is time-independent and $\min(R_{\rm sep})=0.7557$,
$\max(R_{\rm sep})=0.995$.

Fig. 4 shows $R_{\rm ent}$ at $(\omega_1-\omega_2)t=\pi$ as a function of
$\sigma_A$ and $\sigma_B$. Fig. 5 is a slice of Fig. 4 for $\sigma_B=-1.1\pi$.
Fig. 6 shows the time variation of the two ratios, $R_{\rm sep}$ (line of
circles) and $R_{\rm ent}$ (continuous line), for $\sigma_A=0.98\pi$ and
$\sigma_B=-1.1\pi$.

The results show that both classically and quantum mechanically correlated
photons induce correlations on the distant electron interference experiments.
We have compared and contrasted two examples: the $\rho_{\rm sep}$ of Eq.
(\ref{rho_sep}), which is a mixed state; and the $\rho_{\rm ent}$ of Eq.
(\ref{rho_ent}), which is a maximally entangled pure state. These two density
matrices of the electromagnetic field differ only by off-diagonal elements. We
have shown that the effect of these off-diagonal elements on the correlations
between the electron interference experiments, is drastic (compare and
contrast Figs. 3 and 4).

\section{Discussion}
We have considered electron interference experiments irradiated with
nonclassical electromagnetic fields. In this case the phase factor is the
quantum mechanical operator of Eq.(\ref{q_definition}) and its expectation
value with respect to the density matrix of the electromagnetic field, affects
the interference. In this general context, we have studied the case of two
electron interference experiments that are far from each other and are
irradiated with two electromagnetic fields of frequencies $\omega_1,\omega_2$.
The two electromagnetic fields are produced by the same source and are
correlated; consequently the expectation values of the two phase factor
operators in the two experiments are also correlated.

The examples of Eqs. (\ref{rho_sep}) and (\ref{rho_ent}) have been considered.
They represent classically correlated (separable) and quantum mechanically
correlated (entangled) electromagnetic fields, correspondingly. Due to the
correlations in the electromagnetic field the electron fringes are also
correlated. This has been quantified with the ratio $R$ of Eq. (\ref{ratio}).
In the example considered the $R_{\rm sep}$ (Eq. (\ref{R_sep})) is
time-independent and takes values less than $1$. The $R_{\rm ent}$ (Eq.
(\ref{R_ent})) oscillates sinusoidally in time (with frequency $\omega_1 -
\omega_2$) around the value $R_{\rm sep}$. In the case $\omega_1 =\omega_2$
the ratio $R_{\rm ent}(\sigma_A,\sigma_B)$ differs from  $R_{\rm
sep}(\sigma_A,\sigma_B)$ by a constant (which depends on $\sigma _A$,
$\sigma_B$).

Other examples, can also be calculated. But the examples considered show
clearly the main point of the paper, which is that distant electron
interference experiments can be correlated through correlated photons. We have
also shown that the correlations of these distant electron interference
fringes are sensitive to the off-diagonal elements of the electromagnetic
density matrix.

The work brings together concepts from generalized Aharonov-Bohm phenomena
irradiated with nonclassical electromagnetic fields and concepts from
nonclassical correlations and entanglement. The results demonstrate that
entangled electromagnetic fields interacting with electrons produce entangled
electrons.

\begin{figure}[htbp]
\begin{center}
\scalebox{0.43}{\includegraphics{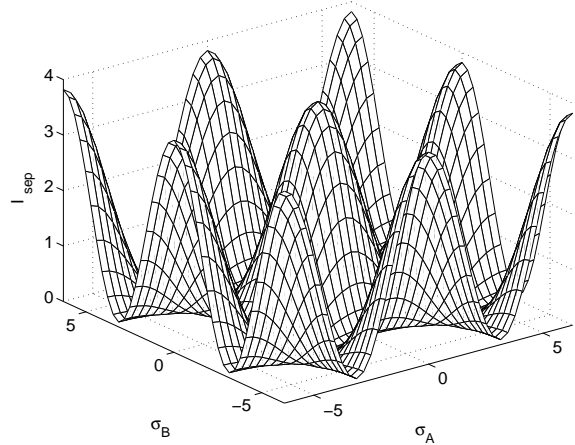}}
\end{center}
\caption{$I_{\rm sep}$ as a function of $\sigma_A,\sigma_B \in
[-2\pi,2\pi]$. The frequencies are $\omega_1=1.2\times
10^{-4}$ and $\omega_2=10^{-4}$, in units where $k_B=\hbar=c=1$.}
\end{figure}
\begin{figure}[htbp]
\begin{center}
\scalebox{0.43}{\includegraphics{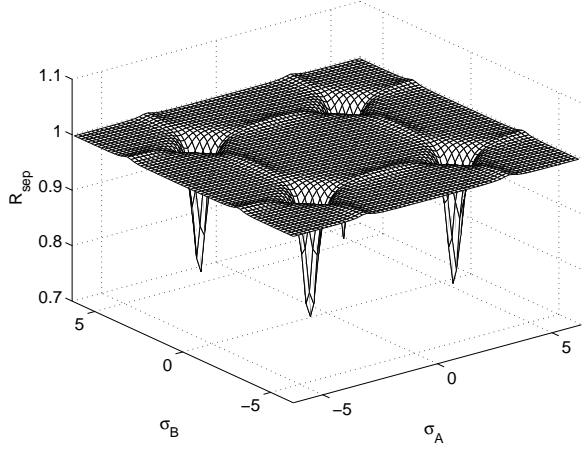}}
\end{center}
\caption{$R_{\rm sep}$ as a function of $\sigma_A,\sigma_B \in
[-2\pi,2\pi]$. Here $\min(R_{\rm sep})=0.7557$ and $\max(R_{\rm sep})=0.995$.
The frequencies are $\omega_1=1.2\times
10^{-4}$ and $\omega_2=10^{-4}$, in units where $k_B=\hbar=c=1$.}
\end{figure}
\begin{figure}[htbp]
\begin{center}
\scalebox{0.43}{\includegraphics{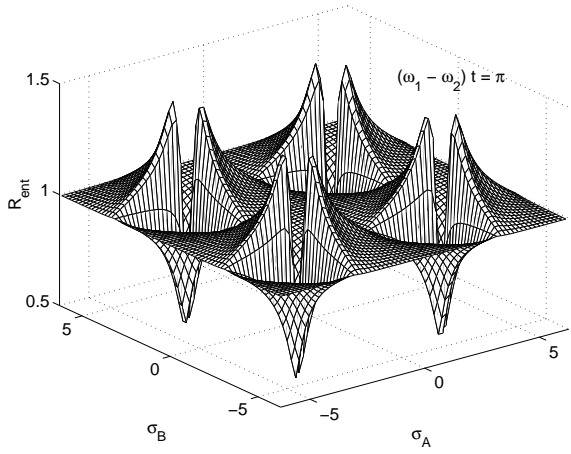}}
\end{center}
\caption{$R_{\rm ent}$ as a function of $\sigma_A,\sigma_B \in
[-2\pi,2\pi]$, at $t=(\omega_1-\omega_2)^{-1}\pi$. The
frequencies are $\omega_1=1.2\times 10^{-4}$ and
$\omega_2=10^{-4}$, in units where $k_B=\hbar=c=1$.}
\end{figure}
\begin{figure}[htbp]
\begin{center}
\scalebox{0.43}{\includegraphics{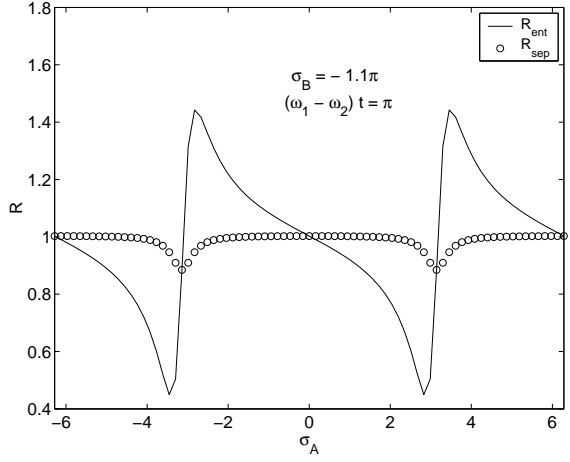}}
\end{center}
\caption{Comparison of $R_{\rm ent}$ (continuous line) and $R_{\rm
sep}$ (line of circles) against $\sigma_A$ for
$(\omega_1-\omega_2)t=\pi$ and $\sigma_B=-1.1\pi$. Note that $\max(R_{\rm sep})=0.995$.
The frequencies are $\omega_1=1.2\times 10^{-4}$ and
$\omega_2=10^{-4}$, in units where $k_B=\hbar=c=1$.}
\end{figure}
\begin{figure}[htbp]
\begin{center}
\scalebox{0.43}{\includegraphics{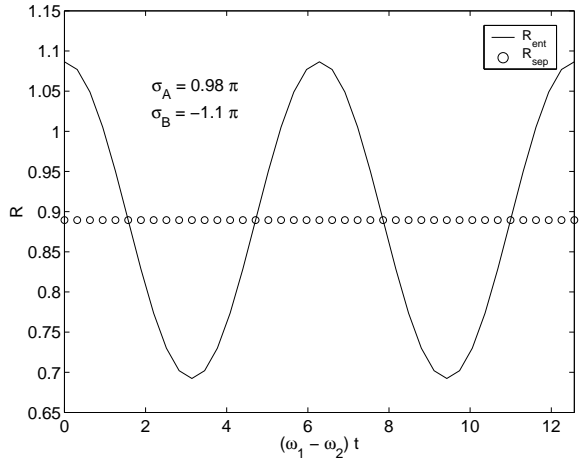}}
\end{center}
\caption{Comparison of $R_{\rm ent}$ (continuous line) and $R_{\rm
sep}$ (line of circles) for $\sigma_A=0.98\pi$ and
$\sigma_B=-1.1\pi$ as a function of dimensionless time. The
frequencies are $\omega_1=1.2\times 10^{-4}$ and $\omega_2=10^{-4}$. We
use units where $k_B=\hbar=c=1$.}
\end{figure}


\end{document}